\documentclass[preprint,prl,10pt,letterpaper,twocolumn]{revtex4}%
\usepackage{amsfonts}
\usepackage{amsmath}
\usepackage{amssymb}
\usepackage{graphicx}%
\providecommand{\U}[1]{\protect\rule{.1in}{.1in}}

\begin{document}
\title{Quantum behavior of a many photons cavity field revealed by quantum discord}
\author{D. Z. Rossatto}
\affiliation{Departamento de F\'{\i}sica, Universidade Federal de S\~{a}o Carlos, P.O. Box
676, 13565-905, S\~{a}o Carlos\textit{, }SP\textit{, }Brazil}
\author{T. Werlang}
\affiliation{Departamento de F\'{\i}sica, Universidade Federal de S\~{a}o Carlos, P.O. Box
676, 13565-905, S\~{a}o Carlos\textit{, }SP\textit{, }Brazil}
\author{E. I. Duzzioni}
\affiliation{Instituto de F\'{\i}sica, Universidade Federal de Uberl\^{a}ndia, Av. Jo\~{a}o
Naves de \'{A}vila, 2121, Santa M\^{o}nica, 38400-902, Uberl\^{a}ndia, MG, Brazil}
\author{C. J. Villas-Boas}
\email{celsovb@df.ufscar.br}
\affiliation{Departamento de F\'{\i}sica, Universidade Federal de S\~{a}o Carlos, P.O. Box
676, 13565-905, S\~{a}o Carlos\textit{, }SP\textit{, }Brazil}
\begin{abstract}
We investigate the quantum-to-classical\ crossover of a dissipative optical
cavity mode based on measurement of the correlations between two atoms which
do not interact with each other, but interact with the cavity mode. Firstly,
we note that there is a time window where the mode has a classical behavior,
which depends on the cavity decay rate, the atom-field coupling strength and
the number of atoms. Then, considering only two atoms inside the cavity and
working in the steady state of the system, we note that the entanglement
between the atoms disappears with the increasing the mean number of photons of
the cavity field ($\overline{n}$). However, the quantum discord reaches an
asymptotic non-zero value, even in the limit of $\overline{n}\rightarrow
\infty$. This happens either by increasing $\overline{n}$ coherently (applying
a coherent driving field) or\ incoherently (raising the temperature of the
reservoir coupled to the cavity mode). Therefore, the cavity mode, which is
quantum by construction, always preserves its quantum behavior in the
asymptotic limit and this is revealed only by the quantum discord.

\end{abstract}

\maketitle

Although the quantum theory predicts many non-classical and intriguing
phenomena as quantum superposition of states and quantum nonlocality
\cite{EPR, ASPECT}, such phenomena can hardly be observed in the macroscopic
world, being the classical physics recovered from the quantum theory for large
excitation numbers and many particle systems \cite{bohr1976}. The emergence of
classical physics from quantum mechanics is therefore actively studied and
suggestions to explain it include decoherence due to\ the interaction with the
environment \cite{schlosshauer2007}, impossibility of macroscopic
superposition of distinct states \cite{ghirardi1986}, and restrictions due to
imprecise measurements \cite{kofler2007}. However, to verify the quantum or
classical behavior of a given system we need to introduce the meter to observe
its properties, which in quantum theory is not a simple task. The simple
interaction between the system and the meter modifies its dynamics in a way
that the quantum-to-classical crossover depends on the meter involved. For
cavity fields, e.g., in microwave \cite{haroche} or circuit \cite{circuit}
systems, usually one employs a single atom as the meter for the cavity field
properties, as for example in Refs. \cite{everit, fink2010, piza}. In Ref.
\cite{fink2010} the quantum-to-classical crossover was investigated by raising
gradually the effective temperature of a coplanar transmission line resonator
strongly coupled to a superconducting artificial atom, i.e., a circuit QED
system. At low temperatures they could observe vacuum Rabi oscillations and
mode splitting, revealing the quantum nature of the light field. However,
these effects disappear when one raises the effective temperature, i.e.,
increasing the mean number of photons of the cavity mode. Naturally, the
conditions needed for a bosonic mode to have a classical behavior depends on
the system parameters and even a cavity mode with a very small mean number of
photons may behave classically \cite{piza}. But, usually, the increasing of
the mean number of photons of the cavity field kills its quantum properties,
as shown in Refs. \cite{everit, fink2010, piza}. However, in these works this
was observed using only one atom interacting with the cavity mode so that the
question about what happens when we use a different meter is still open. Here
we try to answer this question, investigating the behavior of a dissipative
cavity field interacting with $N$ atoms instead of just one. As in \cite{piza}
here we also assume that the cavity mode is pumped by a classical (external)
field which controls the mean number of photons inside the cavity. It can be
easily shown that a purely classical field is not able to generate any kind of
quantum correlation between the atoms. Then, when the cavity mode behaves
classically, all the quantum correlations must be null. Here these quantum
correlations are quantified by quantum discord \cite{zurek2001} and
entanglement of formation \cite{EoF}. On the other hand, the quantum
correlations can be non-zero only when the cavity mode has a quantum behavior,
owing of the indistinguishability of the photons inside the cavity. But, once
some quantum states of the cavity mode are not able to generate quantum
correlations between the atoms, the absence of quantum correlations does not
give us information about the character of the cavity field. In this way, the
presence of quantum correlations between the atoms works out as a signature of
the\textit{ non-classical} behavior of the cavity field.

Considering a cavity mode interacting with $N$ identical two-level atoms
($\left\vert g\right\rangle $ = \ ground state, $\left\vert e\right\rangle $ =
\ excited state) and simultaneously driven by a classical field, the total
Hamiltonian which describes such a system is ($\hslash=1$)%
\begin{equation}
H=\frac{\omega_{0}}{2}S_{z}+\omega a^{\dag}a+H_{P}+H_{I}, \label{Htotal}%
\end{equation}
where $S_{z}=\Sigma_{j=1}^{N}\sigma_{z}^{j}$, being $\omega_{0}$ and
$\sigma_{z}^{j}$ ($=\left\vert e\right\rangle _{j}\left\langle e\right\vert
-\left\vert g\right\rangle _{j}\left\langle g\right\vert $) the atomic
transition frequency and the $z$-Pauli matrix of the atom $j$, respectively;
$\omega$ is the cavity mode frequency and $a$ ($a^{\dagger}$) its annihilation
(creation) operator; $H_{P}=\varepsilon\left(  ae^{i\omega_{L}t}+h.c.\right)
$ describes the pumping field on the cavity mode, $\varepsilon$ and
$\omega_{L}$ being the\ strength and frequency of the driving field,
respectively. Here $h.c.$ stands for hermitean conjugate. $H_{I}=g\sqrt
{N}\left(  aS_{+}+h.c.\right)  $ is the interaction Hamiltonian between the
cavity mode and atoms, with $S_{+}=S_{-}^{\dag}=\frac{1}{\sqrt{N}}\Sigma
_{j=1}^{N}\sigma_{+}^{j}$ , $\sigma_{+}^{j}=\left\vert e\right\rangle
_{j}\left\langle g\right\vert $, and $g$ the atom-field coupling.Writing the
Hamiltonian in a rotating frame with the laser frequency through the unitary
transformation $T=\exp\left[  -i\omega_{L}t\left(  a^{\dag}a+S_{z}/2\right)
\right]  $ we have
\begin{equation}
V_{L}\left(  t\right)  =\delta a^{\dag}a+\frac{1}{2}\Delta S_{z}+\left(
g\sqrt{N}S_{+}a+\varepsilon a+h.c.\right)  , \label{VL}%
\end{equation}
being $\Delta=\omega_{0}-$ $\omega_{L}$ and $\delta=\omega-\omega_{L}$.
Assuming a leaking cavity, the dynamics of this system is governed by the
master equation ($T=0K$)
\begin{equation}
\dot{\rho}=-i\left[  V_{L}\left(  t\right)  ,\rho\right]  +\kappa
\mathcal{L}\left[  a\right]  \rho, \label{eqmestraint}%
\end{equation}
where $\kappa$ is the dissipation rate of the cavity mode and $\mathcal{L}%
\left[  A\right]  \rho=\left(  2A\rho A^{\dag}-A^{\dag}A\rho-\rho A^{\dag
}A\right)  $. As in \cite{everit}, we neglect the atomic decay as the atoms
work out as a meter to monitor the behavior of the cavity mode. To observe the
action of the driven cavity field on the atoms, firstly we apply a
time-independent unitary transformation which consists of a displacement
operation $D\left(  \alpha\right)  =e^{\alpha a^{\dagger}-\alpha^{\ast}a}$,
i.e., $\tilde{\rho}=D^{\dag}\left(  \alpha\right)  \rho D\left(
\alpha\right)  $. Imposing $\alpha=-i\varepsilon/\left(  \kappa+i\delta
\right)  $, we finally obtain%

\begin{equation}
\frac{d\tilde{\rho}}{dt}=-i\left[  H_{JC}+H_{SC},\tilde{\rho}\right]
+\kappa\mathcal{L}\left[  a\right]  \tilde{\rho}, \label{eqmestratil}%
\end{equation}
with $H_{JC}=\delta a^{\dag}a+g\sqrt{N}\left(  aS_{+}+h.c.\right)  ,$
$H_{SC}=\frac{1}{2}\Delta S_{z}+\left(  \Omega S_{+}+h.c.\right)  $ and
$\Omega=g\sqrt{N}\alpha$. We must observe that the chosen value for $\alpha$
is exactly the amplitude of the asymptotic coherent field of the cavity mode
for $\varepsilon\gg g\sqrt{N}$. Under this condition we verify, in all
numerical calculations we have done bellow, the cavity field presents the
statistical properties of a coherent field (i.e., correlation function
$g^{\left(  2\right)  }\left(  0\right)  \,=1$, Mandel factor $Q=0$, and mean
number of photons $\overline{n}=\left\vert \alpha\right\vert ^{2}$).\ Looking
at the atoms, it is clear from Eq. (\ref{eqmestratil}) that, in this displaced
picture, we have two kinds of dynamics: one governed by a classical field and
another one governed by a quantum field. Now we proceed to investigate what
happens to the dynamics of the $N$ two-level atoms when the cavity mode
dissipates strongly.

Firstly we consider $\delta=\Delta_{j}=0$ and assume the weak coupling limit
such that $\kappa\gg g_{eff}\sqrt{\bar{n}_{D}+1}$, with $g_{eff}=g\sqrt{N}$
and\ $\bar{n}_{D}$ the mean number of photons in the cavity mode in the
displaced representation. For $t\gg1/\kappa$ \textbf{ }we can do an adiabatic
elimination of the field variables \cite{livro-zoller}, resulting in a reduced
master equation for the atoms, which in the interaction picture is given by
\begin{equation}
\dot{\rho}_{a}=-i\left[  H_{SC},\rho_{a}\right]  +\Gamma_{eff}\mathcal{L}%
\left[  S_{-}\right]  \rho_{a}, \label{eqmestraatomica}%
\end{equation}
with $H_{SC}=\left(  \Omega_{eff}S_{+}+h.c.\right)  $, $\Omega_{eff}%
=g\alpha\sqrt{N}$, and $\Gamma_{eff}=g^{2}N/\kappa$.

We have to notice that (\ref{eqmestraatomica}) describes a set of $N$ atoms
driven by a classical field with effective Rabi frequency $\left\vert
\Omega_{eff}\right\vert $ and interacting with a common effective reservoir
with an effective decay rate $\Gamma_{eff}$. For $\left\vert \Omega
_{eff}\right\vert \gg\Gamma_{eff}$ , i.e., for $\varepsilon\gg g\sqrt{N}$, and
for interaction time $t\ll1/\Gamma_{eff}=\kappa/g^{2}N$, according to Eq.
(\ref{eqmestraatomica}) we can see that the dynamics of the system will be
governed mainly by a free evolution. Then, as the effective master equation
above was derived for $t\gg1/\kappa$, we can see that, for interaction times
limited to the time window%
\begin{equation}
1\ll\kappa t\ll\frac{\kappa}{\kappa_{eff}}=\left(  \frac{\kappa}{g\sqrt{N}%
}\right)  ^{2}, \label{janela}%
\end{equation}
the effective master equation can be approximated by $\dot{\rho}_{a}%
\simeq-i\left[  H_{SC},\rho_{a}\right]  $,\ which represents an atomic system
interacting with a classical electromagnetic field. Starting in a separable
atomic state, the atomic system remains separable, i.e., the interaction of
the atoms with a common classical field is not able to generate any kind of
correlations between the atoms. Moreover, if an initial atomic state is pure,
it will remain pure for any time. In this way, the atomic purity is a good
parameter to visualize the validity this semi-classical approximation, once
out of the time window (\ref{janela}) the dynamics of the system is governed
by Eq. (\ref{eqmestraatomica}) where the presence of the effective dissipative
term introduces decoherence in the atomic system.

Looking at the time window we see that, for a fixed $g$, the interaction time
where the semi-classical regime is still valid decreases with $1/N$, i.e., as
bigger the number of atoms inside the cavity, the smaller the time window in
which the semi-classical regime is valid. In Fig. \ref{purity_qd_eof}(a) we
show the atomic purity for 1, 2 and 3 atoms inside the cavity, assuming
$g=0.01\kappa$, $\varepsilon=1.0\kappa$ (which results in a maximum mean
number of photons $\overline{n}_{\max}=\left\vert \varepsilon/\kappa
\right\vert ^{2}=1$)$,$ and considering the atoms prepared initially in the
excited state $\left\vert e\right\rangle $. For atoms initially prepared in
the\ ground state $\left\vert g\right\rangle $ the graphic is qualitatively
the same. We can see in this figure that the purity of the system decreases
quickly when we go out of the time window which defines the semi-classical
regime.\textbf{
\begin{figure}
[t]
\begin{center}
\includegraphics[
trim=0.208382in -0.052070in -0.208382in 0.052070in,
height=1.6302in,
width=3.3589in
]%
{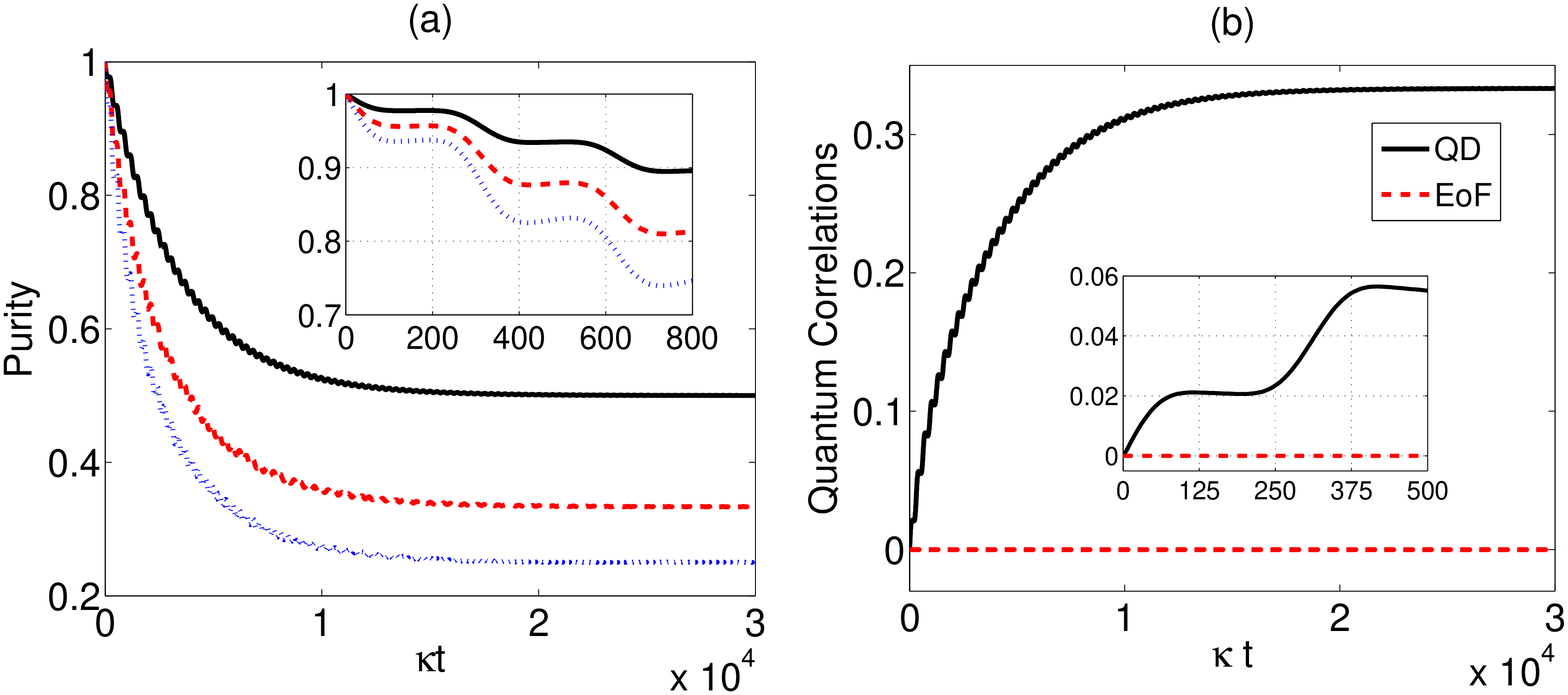}%
\caption{Time evolution of the (a) atomic purity for one (full line), two
(dashed) and three (dotted) atoms within the cavity and (b) quantum
correlations for two atoms: QD (full line) and EoF (dashed). We consider
$g=0.01\kappa$, $\varepsilon=\kappa$, and all atoms initially in the excited
state $\left\vert e\right\rangle \,$. The EoF is always zero while quantum
discord grows until it reaches the value $\simeq0.33$ asymptotically,
revealing the quantum nature of the field for long interaction times.}%
\label{purity_qd_eof}%
\end{center}
\end{figure}
}

The next step consists in determining the quantum correlations between two
atoms (here $A$ and $B$). The measure of total quantum correlations used is
the \textit{quantum discord} (QD) \cite{zurek2001}. Nonzero QD in a bipartite
system implies that it is impossible to extract all information about one
subsystem without perturbing its complement. In every cases studies here the
reduced density matrix for the atomic system $\rho_{AB}$ has the X structure
defined by its elements $\rho_{12}=\rho_{13}=\rho_{24}=\rho_{34}=0$, with real
coherences and $\rho_{22}=\rho_{33}$. For this class of density matrix the QD
can be analytically calculated \cite{nonmarkov}: $QD(\rho_{AB})=S(\rho
_{A})-S(\rho_{AB})-\max\left\{  D_{1},D_{2}\right\}  $, where $D_{1}=%
{\displaystyle\sum\limits_{i=1,3}}
\rho_{ii}\log_{2}\left(  \frac{\rho_{ii}}{\rho_{ii}+\rho_{i+1,i+1}}\right)  +%
{\displaystyle\sum\limits_{i=2,4}}
\rho_{ii}\log_{2}\left(  \frac{\rho_{ii}}{\rho_{ii}+\rho_{i-1,i-1}}\right)  $
and $D_{2}=%
{\displaystyle\sum\limits_{i=0,1}}
\left(  \frac{1+\left(  -1\right)  ^{k}\beta}{2}\right)  \log_{2}\left(
\frac{1+\left(  -1\right)  ^{k}\beta}{2}\right)  ,$ with $\beta^{2}=(\rho
_{11}-\rho_{44})^{2}+4(\left\vert \rho_{23}\right\vert +\left\vert \rho
_{14}\right\vert )^{2}$. Here $S(\cdot)$ denotes the von Neumann entropy
\cite{nielsen} and $\rho_{A}=\mbox{Tr}_{B}\rho_{AB}$. The entanglement,
another kind of quantum correlation, is computed through \textit{entanglement
of formation} (EoF) \cite{EoF}. For X form density matrix the EoF is
\cite{nonmarkov,wootters} $EoF(\rho_{AB})=-\eta\log_{2}\eta-(1-\eta)\log
_{2}(1-\eta),$ where $\eta=\frac{1}{2}\left(  1+\sqrt{1-C^{2}}\right)  $ with
$C=2\max\left\{  0,|\rho_{14}|-\sqrt{\rho_{22}\rho_{33}},|\rho_{23}%
|-\sqrt{\rho_{11}\rho_{44}}\right\}  $ being the concurrence. Although for
pure states the QD is equal to EoF, this is not the case for mixed states,
existing quantum correlated states with null entanglement.

Now, for two atoms initially prepared in the state $\left\vert
e,e\right\rangle $, we see that the EoF is zero all the time (for initial
state $\left\vert e,g\right\rangle $ or $\left\vert g,g\right\rangle $ the EoF
is null for long interaction times and almost zero for very short interaction
times), as we see in Fig. \ref{purity_qd_eof}(b). In this way, the EoF is not
useful to distinguish the quantum and classical character of a cavity field in
the steady state of the system. However, the QD has very small values within
the time window (\ref{janela}), so that the correlations generated by the
cavity field in the atoms are negligible, which confirms the classical
character of the field within this window. Moreover, the QD grows continuously
until it reaches a stationary value asymptotically. This considerable value
($\simeq0.33$) for the QD for initial state $\left\vert e,e\right\rangle $ (or
$\left\vert g,g\right\rangle $) shows that the quantum correlations between
atoms, generated via interaction of the atoms with the cavity mode, is
significant for long interaction times and reveals the quantum nature of this
cavity field. Thus, to determine the classical or quantum behavior of the
field, we must calculate the correlations between atoms in the steady state.

\begin{figure}
[t]
\begin{center}
\includegraphics[
trim=0.942504in 0.034739in -0.942504in -0.034739in,
height=2.2139in,
width=4.4287in
]%
{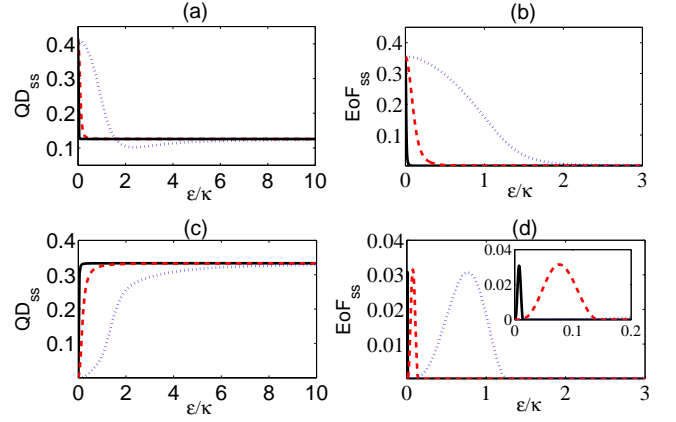}%
\caption{Stationary quantum correlations versus $\varepsilon/\kappa$ for
initial atomic states $\left\vert e,g\right\rangle $ (a and b) and $\left\vert
g,g\right\rangle $ (c and d). The atom-field coupling was fixed as
$g=0.01\kappa$, (full line), $0.1\kappa$ (dashed line), and $1.0\kappa$
(dotted line). For $\varepsilon\gg g$ the EoF (b and d) goes to zero while QD
goes to the finite value $\simeq0.12$ (a) or $\simeq0.33$ (c). }%
\label{d_eofvse_variando_g_2}%
\end{center}
\end{figure}

\textit{Entanglement and Quantum Discord in the stationary regime: }here we
analyze the stationary behavior of the QD and EoF as function of the ratios
$g/\kappa$ and $\varepsilon/\kappa$ through the numerical solution of the Eq.
(\ref{eqmestraint}) without any approximation. In Figs.
\ref{d_eofvse_variando_g_2}(a) and (b), for initial atomic state $\left\vert
e,g\right\rangle $ (or $\left\vert g,e\right\rangle $), and
\ref{d_eofvse_variando_g_2}(c) and (d), for initial atomic state $\left\vert
g,g\right\rangle $ (or $\left\vert e,e\right\rangle $), we see that EoF always
goes to zero for $\varepsilon\gg g$, being different of zero only for
$\varepsilon\ll g$, i.e., for small mean number of photons. These results are
in accordance with the equivalence principle since for very high mean number
of photons we expect an agreement between quantum and semi-classical
description, which means no quantum correlations between the atoms. But,
surprisingly, the QD is always different of zero, reaching a significative
value in the limit of $\varepsilon\gg g$ ($QD_{ss}\simeq0,12$ for initial
state $\left\vert e,g\right\rangle $ and $QD_{ss}\simeq0,33$ for initial state
$\left\vert g,g\right\rangle $). This means that the cavity mode is able to
generate quantum correlations between the atoms for any finite mean number of
photons inside the cavity, even for extremely intense fields. In this way the
cavity mode, which is quantum by construction, has its quantum character
revealed only through the QD between the atoms.

\begin{figure}
[t]
\begin{center}
\includegraphics[
trim=0.160142in 0.074294in -0.160142in -0.074294in,
height=1.4736in,
width=3.2993in
]%
{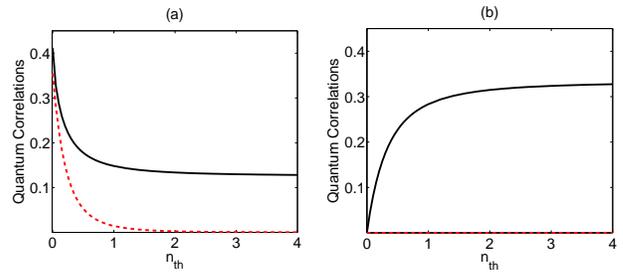}%
\caption{Stationary quantum correlations versus mean number of thermal photons
of the reservoir: QD (full line) and EoF (dashed line). We have considered
atom-field coupling $g=0.01\kappa$ and initial atomic state (a) $\left\vert
e,g\right\rangle $ and (b) $\left\vert gg\right\rangle $. In (b) the EoF is
zero for all $n_{th}$. }%
\label{quantum_correlations_ss_temperature}%
\end{center}
\end{figure}

The origin of the quantum character is the indistinguishability of paths in
the exchange of photons between atoms and field. As the photon is indivisible,
when a photon is absorbed from the field it generates a superposition of
possibilities: either the photon is absorbed by the first atom or is absorbed
by the second one, but these two possibilities happen simultaneously. One
could argue that the origin of this quantum character could be in the
coherence of the driving field, which generates a coherent field inside the
cavity as in \cite{alsing}. However, being this the case, an incoherent
pumping of photons into the cavity would not generate correlations between the
atoms in the stationary regime. To analyze this point more carefully, we have
neglected the driving field and assumed that the cavity mode is at a finite
temperature $T$, which implies in an incoherent injection of photons into the
cavity. In this case we have to consider $\varepsilon=\delta=0$ and replace
the master equation (\ref{eqmestraint}) by%
\[
\dot{\rho}=-i\left[  V_{L}\left(  t\right)  ,\rho\right]  +\kappa\left(
n_{th}+1\right)  \mathcal{L}\left[  a\right]  \rho+\kappa n_{th}%
\mathcal{L}\left[  a^{\dag}\right]  \rho,
\]
with $n_{th}$ being the mean number of thermal photons. For one atom
interacting with a cavity mode and in the limit of large photon numbers, the
classical limit derived in \cite{fink2010} requires $n_{th}>\left(
g/\kappa\right)  ^{2}$. In Fig. \ref{quantum_correlations_ss_temperature} we
have assumed $g=0.1\kappa$ which implies that $n_{th}=1$ is already much
bigger than $\left(  g/\kappa\right)  ^{2}$. We see that the EoF goes to zero
in the steady state as increase the temperature. However, this does not happen
to the QD, so that the cavity mode is still able to generate quantum
correlations between the atoms even for a cavity mode interacting with a
thermal reservoir, as we can see in Figs.
\ref{quantum_correlations_ss_temperature}(a) and (b). Taking in to account the
initial atomic state $\left\vert e,g\right\rangle $ we note that the QD and
the EoF decay when we increase $n_{th}$. The EoF goes to zero while the QD
reaches the asymptotic value $\simeq0.12$ as in the case we have a coherent
injection of photons. For the initial atomic state $\left\vert
g,g\right\rangle $ we see that the EoF is always zero and for low temperatures
the QD is negligible. But, increasing $n_{th}$, the QD increases, reaching the
asymptotic value $QD_{ss}\simeq0.33$, as when we have a coherent driving
field. In this case exactly the same behavior of quantum correlations is
obtained for any value of atom-field coupling. This happens due the
thermalization of the system, i.e., the cavity mode thermalizes with the
reservoir and the atoms effectively thermalize with the cavity mode. The
atom-field coupling just determines the interaction time required for the
thermalization of the atoms with the cavity field. Therefore, the stronger
this coupling, the shorter the interaction time required to reach the
stationary state of the system. Thus we see that the cavity field is always
able of generating quantum correlations, no matter the temperature of the
reservoir, revealing the quantum character of the field for any temperature.

As a conclusion, we showed that a non-zero QD between atoms works out as a
signature of the non-classical behavior of the cavity mode. It is important to
notice that, for high mean number of photons inside the cavity, other quantum
correlations such as entanglement is not present in the atomic system such
that the quantum character of the cavity mode can be revealed only by QD.
Also, the QD is different of zero for any value of atom-field coupling and
even in the limit of very strong driving field or high temperatures, both
implying in many photons inside the cavity, revealing us the quantum nature of
the cavity mode even in these limits. (The atom-field coupling is important to
determine the interaction time required to reach the stationary state of the
system.) Then, we can conclude that the QD between the atoms show us that
classical behavior of the cavity field is not recovered in the limit of many
photons, contrary to the correspondence principle. Moreover, there is a time
window where the mode has an effective classical behavior, which depends on
the cavity decay rate, the strength of the atom-field coupling and the number
of atoms.

\begin{acknowledgments}
The authors would like to thank the Brazilian agency CNPq and the Brazilian
National Institute of Science and Technology for Quantum Information (INCT-IQ).
\end{acknowledgments}

\end{document}